\documentclass[11pt,a4paper]{article}
\usepackage{jheppub}

\usepackage{amsmath}
\usepackage{latexsym}

\title{\Large{Causality in noncommutative spacetime}}

\author{Everton M. C. Abreu$^{a,b,c}$}
\author{and Mario J. Neves$^a$}

\affiliation{$^a$Grupo de F\' isica Te\'orica, Departamento de F\'{\i}sica, \\
Universidade Federal Rural do Rio de Janeiro\\
BR 465-07, 23890-971, Serop\'edica, Rio de Janeiro, Brazil \\\\ 
$^b$Centro Brasileiro de Pesquisas F\' isicas (CBPF),\\
Rua Dr. Xavier Sigaud 150, Urca, \\
CEP  22290-180, Rio de Janeiro, Brazil\\\\
${}^{c}$Departamento de F\'{\i}sica, ICE, Universidade Federal de Juiz de Fora,\\
36036-330, Juiz de Fora, MG, Brazil\\\\
\bigskip
\today\\}
\emailAdd{evertonabreu@ufrrj.br, mariojr@ufrrj.br}

\abstract{In this paper we investigated the causality problem present in the recent work about the Doplicher-Fredenhagen-Roberts-Amorim (DFRA) noncommutative framework which analyzed the complex scalar field.  
To accomplish this task we provided a brief review of the main ingredients of the problem and we demonstrated precisely that the DFRA algebra obeys the rules of the Canonical Commutation Relations algebra.  This fact permitted us to prove the form of the DFRA operators previously constructed in the usual way. 
After that, we introduced the solution of its Klein-Gordon equation with a source term.  Its solution was accomplished through the retarded, advanced and causal Green functions constructed in this noncommutative ten dimensional DFRA spacetime.  We believe that this solution constitutes the first step in the elaboration of a quantum field theory using the DFRA formalism where the noncommutative parameter is an ordinary coordinate of the system and therefore has a canonical conjugate momentum.}


\usepackage{color}
\usepackage{latexsym}
\usepackage{amsmath}
\usepackage{amssymb}
\usepackage{eufrak}
\usepackage{euscript}
\usepackage{pstricks}
\usepackage{graphics}
\usepackage{graphicx}
\usepackage{picture}

\newcommand{\be}{\begin{equation}}
\newcommand{\ee}{\end{equation}}
\newcommand{\ba}{\begin{eqnarray}}
\newcommand{\ea}{\end{eqnarray}}
\newcommand{\p}{\partial}
\def\ni{\noindent} 

\def\no{\nonumber}

\begin{document}

\maketitle

\pagestyle{myheadings}
\markright{Causality in noncommutative spacetime}


\section{Introduction}
\renewcommand{\theequation}{1.\arabic{equation}}
\setcounter{equation}{0}

The inconvenience of having infinities that destroy the final results of several calculations in quantum field theory motivated theoretical physics to ask if a continuum spacetime would be really necessary.  The obvious alternative would be to construct a discrete spacetime with a noncommutative (NC) algebra.

Grabbing these ideas Snyder \cite{Snyder} published a first work considering the spacetime as being NC.  However, Yang \cite{yang} demonstrated that Snyder's hopes about getting rid of the infinities were not achieved by noncommutativity.  This demonstration doomed the Snyder NC theory to years of ostracism.
After the important result that the algebra obtained with a string theory embedded in a magnetic background is NC, a new flame concerning noncommutativity was rekindle \cite{sw,bg}.

One of the alternatives (the most famous at least) of introducing noncommutativity is through the Moyal-Weyl product where the NC parameter ($\theta^{\mu\nu}$) is an antissimetric constant \cite{QG}.  However, at superior orders of calculations the Moyal-Weyl product turns out to be highly nonlocal.  This fact forced us to work with low orders in $\theta^{\mu\nu}$.  

One and more recent way to work with noncommutativity was introduced by Doplicher-Fredenhagen-Roberts (DFR) which considered $\theta^{\mu\nu}$ as an ordinary coordinate of the system \cite{DFR}.  Recently, in \cite{Amorim1} Amorim came across with the idea of constructing an extension of the DFR spacetime introducing the conjugate canonical momentum associated with $\theta^{\mu\nu}$ (for a review the reader can see also \cite{amo}).  This extended and new NC spacetime has ten dimensions: four relative to Minkowski spacetime and six relative to $\theta$-space. 

After this first work, a new version of NC quantum mechanics (NCQM) \cite{Amorim1,Amorim4,Amorim5,Amorim2} was introduced, where not only the coordinates ${\mathbf x}^\mu$ and their canonical momenta ${\mathbf p}_\mu$ are considered as operators in a Hilbert space ${\cal H}$, but also the objects of noncommutativity $\theta^{\mu\nu}$ and their canonical conjugate momenta $\pi_{\mu\nu}$.
All these operators belong to the same algebra and have the same hierarchical level, introducing  a minimal canonical
extension of the DFR algebra, the so-called Doplicher-Fredenhagen-Roberts-Amorim (DFRA) formalism. This enlargement of the usual set of Hilbert space operators allows the theory to be invariant under the rotation group $SO(D)$, as showed 
in detail in Ref. \cite{Amorim1,Amorim2}, when the treatment is a nonrelativistic one.  Rotation invariance in a nonrelativistic theory is fundamental if one intends to describe any physical system in a consistent way. 
In Ref. \cite{Amorim4,Amorim5}, the corresponding relativistic treatment was explored, which permits us to implement 
Poincar\'{e} invariance as a dynamical symmetry   \cite{Iorio} in  NCQM  \cite{NCQM}. 
In \cite{aa} it was considered the ``second quantization" of the model discussed in Ref \cite{Amorim4}, showing that the extended Poincar\'e symmetry here is generated via generalized Heisenberg relations, giving the same algebra displayed in \cite{Amorim4,Amorim5}.   However, in our opinion, although some relations between the NC coordinates/momentum and its respective operators is complete, it is still missing a whole and rigorous fathoming of how to construct such operators.

Although it was constructed a NC DFRA Klein-Gordon equation in \cite{aa} with a source term, an effective action using the Green functions was not calculated to the end.  It was only given its general expression.  In other words, here we did explored all the solutions of the KG equation, the homogeneous and non-homogeneous one.  Concerning the non-homogeneous solutions, we can have it with the source term and another solution where the source term is a delta function.  For each one kind, we have a different and well-known approach for the Green function.  However, what is new here is to bring this approach to the NC realm.

In this work we accomplished both tasks computing the advanced, retarded and causal Green functions for the charged scalar field with sources.  And we described the equivalence between the DFRA algebra and the well-known Canonical Commutation Relations (CCR) algebra.  This equivalence helped us to construct precisely the operators involved in the DFRA algebra, instead to follow what the commutators determine, which can cause incorrect constructions.  Namely, so far in the literature, the DFRA operators were constructed only through the usual way.  Here, we provided a rigorous demonstration of the form of these operators.

The subjects are distributed through the paper following a sequential order such as in section II we review Snyder's algebra and its subtleties. In section III we write the mathematical basis of DFRA algebra.  It contains a very brief review of the Canonical Commutation Relations (CCR) algebra.
In section IV we describe the DFRA formalism and we showed the the DFRA algebra obeys the CCR rules.  In section V, to be self-contained, we review the NC Klein-Gordon in DFRA spacetime and describe its symmetries.  In section VI we provide the ingredients for the computation of the NC Green functions.  It was carried out in section VII, where
we computed the retarded and advanced Green functions. The causal Green function (the Feynman propagator) was computed in section VIII and  the conclusions and perspectives for future works are depicted in the ninth section.


\section{Snyder's algebra}
\renewcommand{\theequation}{2.\arabic{equation}}
\setcounter{equation}{0}

The relativity theory can be based on the invariance of the indefinite quadratic form,

\be
\label{A}
s^2\,=\,(x^0)^2\,-\,(x^1)^2\,-\,(x^2)^2\,-\,(x^3)^2\,=\,-\,x_\mu\,x^\mu
\ee

\ni for the transformation of one inertial reference systems to another.

Snyder considered $\hat{\mathbf x}$ as the Hermitian operators for the spacetime coordinates of a particular Lorentz reference system.  He deemed also that $\hat{\mathbf x}$ spacetime coordinates spectra are invariant under Lorentz transformations \cite{bck,kmm}.  We know it is true concerning the standard continuous spacetime.  However, Snyder showed that this solution is not unique.  There is a Lorentz invariant spacetime where there is a natural length unit within.  

The target for Snyder was to find the operators $\hat{\mathbf x}$ that have a Lorentz invariant spectrum.  To accomplish this task, he considered the homogeneous quadratic form

\ba
\label{B}
-\,(y)^2\,&=&\,(y_0)^2\,-\,(y_1)^2\,-\,(y_2)^2\,-\,(y_3)^2\,-\,(y_4)^2 \no \\
&=&-\,y_\mu\,y^\mu\,-\,(y_4)^2\,\,,
\ea

\ni where the $y's$ are real variables.  Hence, $\hat{\mathbf x}$ are defined through the infinitesimal elements of the group under which (\ref{B}) is invariant.  We can write $\hat{\mathbf x}$ as \cite{bck,kmm} 

\be
\label{C}
\hat{\mathbf x}^\mu\,=\,i\,a\,\Big(\,y_4\,\frac{\partial}{\partial y_\mu}\,-\,y^\mu\,\frac{\partial}{\partial y_4}\Big)\,\,,
\ee

\ni where $a$ is the natural length unit.  These mentioned operators work over $y_\mu$ and $y_4$ functions.  The $x^i\:(i=1,2,3)$ spectra are discrete.  But $x^0$ has a continuous spectrum which limit goes from $-\infty$ to $+\infty$.

The transformations that leave invariant $y_4$ and the quadratic form (\ref{B}) are covariant Lorentz transformations cause contravariant Lorentz transformations in $\hat{\mathbf x}$.

Now we will define additional operators,

\be
\label{D}
\hat{\mathbf M}^{\mu\nu}\,=\,i\,a\,\Big(\,y^\mu\,\frac{\partial}{\partial y_\nu}\,-\,y^\nu\,\frac{\partial}{\partial y_\mu}\Big)\,\,,
\ee

\ni which are infinitesimal elements of the four-dimensional Lorentz group.

The ten operators defined in (\ref{C}) and (\ref{D}) have the following commutation relations,

\begin{subequations}
\label{1021}
\ba
\Big[\hat{\mathbf x}^\mu,\hat{\mathbf x}^\nu \Big]\,&=&\,i\,a^2\,\hat{\mathbf M}^{\mu\nu} \,\, \label{1021a} \\
\Big[\,\hat{\mathbf M}^{\mu\nu},\hat{\mathbf x}^\lambda\,\Big]\,&=&\,i\,\Big(\hat{\mathbf x}^\mu\,\eta^{\nu\lambda}\,-\,\hat{\mathbf x}^\nu\,\eta^{\mu\lambda}\Big)\,\, \label{1021b} \\
\Big[\,\hat{\mathbf M}^{\mu\nu},\hat{\mathbf M}^{\alpha\beta}\,\Big]\,&=&\,i\,\Big(\hat{\mathbf M}^{\mu\beta}\,\eta^{\nu\alpha}\,-\,
\hat{\mathbf M}^{\mu\alpha}\,\eta^{\nu\beta}\,+\,\hat{\mathbf M}^{\nu\alpha}\,\eta^{\mu\beta}\,-\,\hat{\mathbf M}^{\nu\beta}\,\eta^{\mu\alpha}\Big)\,\, \label{1021c}
\ea
\end{subequations}

\ni The Lorentz symmetry $SO(3,1)$ given in (\ref{1021c}) was extended to $SO(4,1)$ symmetry determined by Eqs. (\ref{1021}).

We can fathom the position operators $\hat{\mathbf x}^i$, which have discrete spectra, in terms of a position non-zero minimum uncertainty.  It is possible \cite{bck} to obtain the space part of Snyder algebra if we consider it as a generalized Heisenberg algebra.


\section{The CCR algebra}
\renewcommand{\theequation}{3.\arabic{equation}}
\setcounter{equation}{0}

In order to make this paper self-contained, in this section we intend to make a very brief review of the CCR algebra in order to support mathematically the DFRA algebra.  We will show here and in the next section that the commutators defined in DFRA algebra obey the CCR algebra and therefore we can construct some operators of this NC spacetime.  

We will introduce some theorems but we will not provide the respective demonstrations, which can be found in basic references such as \cite{brrs}, which we will follow closely in this section.

With this idea in mind, let us write the Stone's theorem, which states that, if $U(t)$ is a strongly continuous one-parameter unitary group on a Hilbert space ${\cal H}$, then, there is a self-adjoint operator $A$ on ${\cal H}$ so that $U(t)=e^{itA}$.  Demonstration in \cite{brrs}.

Another useful theorem says that if $t \rightarrow U(t)=U(t_1 ,\ldots,t_n )$ is a strong continuous map of $\mathbb{R}^n$ into the unitary operators on a separable Hilbert space ${\cal H}$ satisfying $U(t+s)=U(t)U(s)$ and $U(0)=\mathbb{I}$.  Let $D$ be the set of finite linear combinations of vectors of the form

\be
\label{AAA}
\varphi_f\,=\,\int_{\mathbb{R}^n}\,f(t)U(t)\,\varphi\,dt\qquad,\qquad\varphi\,\in{\cal H}\,\,,f\in\,C^{\infty}_0\,(\mathbb{R}^n)
\ee

\noindent Therefore $D$ is the domain of essential self-adjointness for each of the generators $A_j$ of the one-parameter subgroups $U(0,0,\ldots,t_j ,\ldots,0)$ each $A_j : D \rightarrow D$ and the $A_j$ commutate, $j=1,\ldots,n$.  Furthermore, there is a projection-valued measure $P_{\lambda}$ on $\mathbb{R}^n$ so that

\be
\label{BB}
\Big( \varphi\,,\,U(t)\,\psi\Big)\,=\,\int_{\mathbb{R}^n}\,e^{it\cdot\lambda}\,d(\varphi,P_{\lambda}\,\psi)
\ee

\noindent for all $\varphi, \psi\,\in\,{\cal H}$ \cite{brrs}.

With both theorems and some additional considerations \cite{brrs} we can define that two self-adjoint operators $A$ and $B$ are said to commute if and only if all the projections in theirs associated projection-valued (defined in ($B$)) measures commute.  But notice that if $A$ and $B$ are self-adjoint operators on a Hilbert space ${\cal H}$, then the following three statements are equivalent: (i) $A$ and $B$ commute; (ii) if $Im \lambda$ and $Im \mu$ are nonzero, then 
$R_{\lambda}(A)\,R_{\mu}(B)=R_{\mu}(B)\,R_{\lambda}(A)$ and (iii) for all 
$s,t\,\in \mathbb{R},\,e^{itA}\,e^{isB}\,=\,e^{isB}\,e^{itA}$ \cite{brrs}, where we can write

\ba
\label{CC}
e^{itA} \approx\,\mathbb{I}\,+\,\sum^{\infty}_{n=1}\,\frac{(itA)^n}{n!} \nonumber \\
e^{isB} \approx\,\mathbb{I}\,+\,\sum^{\infty}_{n=1}\,\frac{(isB)^n}{n!}
\ea

So, for example, if a pair $P,Q$ of self-adjoint operators is said to ``obey" the CCR, so we have that

\be
\label{DD}
PQ\,-\,QP\,=\,-\,i\,\mathbb{I}
\ee

\noindent and $P$ and $Q$ cannot both be bounded.  If they were, the relation $PQ^n\,-\,Q^n\,P\,=\,-inQ^{n-1}$ would imply,

\be
\label{E}
n\,\|Q\|^{n-1}\,=\,n\,\|Q^{n-1}\|\,\leq\,2\,\|P\|\,\|Q\|^n
\ee

\noindent and for all $n$, $2\,\|P\|\,\|Q\|\geq\,n$, which is a contradiction.  Hence, either $P$ or $Q$ or both must be unbounded.  The standard realization or ``representation" used in quantum mechanics is Schr\"odinger representation where

\be
\label{E2}
{\cal H}=L^2(\mathbb{R})
\ee

\ni and $P$ and $Q$ are the closures of $i^{-1}d/dx$ and multiplication by $x$ on ${\cal S}(\mathbb{R})$, where ${\cal S}(\mathbb{R})$ is a domain of essential self-adjointness for $i^{-1}d/dx$ and $x$,

\be
\label{F}
{1 \over i}\,\frac{d}{dx}\,:{\cal S}(\mathbb{R}) \rightarrow {\cal S}(\mathbb{R})\qquad,\qquad x:\,{\cal S}(\mathbb{R})\rightarrow {\cal S}(\mathbb{R}) 
\ee

\noindent and for $\varphi\,\in\,{\cal S}(\mathbb{R})$,

\be
\label{G}
{1 \over i}\,\frac{d}{dx}\,(x\varphi)\,-\,x\,\Big({1 \over i}\,\frac{d}{dx}\,\varphi \Big)\,=\,-i\,\varphi\,\,.
\ee
This last considerations will be very important for us the future.

Finally, let us state the so-called von Neumann theorem which says that if $U(t)$ and $V(s)$ are one-parameter, continuous, unitary groups on a separable Hilbert space ${\cal H}$ satisfying the Weyl relations, then, there are closed subspaces ${\cal H}_l$ so that \cite{brrs}: \\

\ni (i) ${\cal H}\,=\,\overset{N}{\underset{l=1}{\bigoplus}}\,{\cal H}_l$ ($N$ a positive integer or $\infty$);\\ 

\ni (ii) $U(t):{\cal H}_l \rightarrow {\cal H}_l ,\,V(s): {\cal H}_l \rightarrow {\cal H}_l$ for all $s,t\,\in\,\mathbb{R}$;\\

\ni (iii) for each $l$ there is a unitary operator $T_l : {\cal H}_l \rightarrow L^2(\mathbb{R})$ such that $T_l \,U(t)\, T^{-1}_l$ is multiplied by $e^{isX}$.\\

We will see the implications of these theorems and definitions in the following sections.  
We will show that the DFRA algebras in fact obey the rules of the CCR where the relations (\ref{DD}) and (\ref{E}) are obeyed by DFRA algebra.  And we can also construct the Schr\"odinger repesentation (\ref{F}) and (\ref{G}) for the DFRA quantum mechanics. 


\section{Quantum Mechanics in the DFRA noncommutative space}
\renewcommand{\theequation}{4.\arabic{equation}}
\setcounter{equation}{0}

These spontaneous manifestations of noncommutativity can lead us, for example, to the fact that standard four-dimensional spacetimes may become NC, namely, that the position four-vector operator ${\mathbf x}^\mu$ obeys the following rule
\be \label{a1}
[\,{\mathbf x}^\mu\,,\,{\mathbf x}^\nu\,]\,=\,i\,\theta^{\mu\nu}\,\,,
\ee

\ni where $\theta^{\mu\nu}$ is a real, antisymmetric and constant matrix.  It can be demonstrated that, in a certain limit, a gauge theory on noncommutative spaces is tantamount to string theory.    

As we said above, in the early works \cite{Snyder,yang}, it was introduced a five dimensional spacetime with $SO(4,1)$ as a symmetry group, with generators ${\mathbf M}^{AB}$, satisfying the Lorentz algebra, where $A,B=0,1,2,3,4$ and using natural units, i.e., $\hbar=c=1$. Moreover, the relation between coordinates and  generators of the $SO(4,1)$ algebra can be written as,
$$\,\,
{\mathbf x}^{\mu}=a\,{\mathbf M}^{4\mu}\,\,
$$
(where $\mu,\nu=0,1,2,3$ and the parameter $a$ has dimension of length), promoting in this way the spacetime coordinates to Hermitian operators. The mentioned relation introduced the commutator,
\begin{equation}
\label{2}
[{\mathbf x}^\mu,{\mathbf x}^\nu] = i a^2{\mathbf M}^{\mu\nu}
\end{equation}
\noindent and the identities (\ref{1021b}) and (\ref{1021c}), which agree with four dimensional Lorentz invariance. 

Some years back, in \cite{DFR}, the authors essentially assumed (\ref{a1}) as well as the vanishing of the triple commutator among the coordinate operators.  The DFR algebra is based on  principles imported from GR and QM. In addition to (\ref{a1}) it also assumed that 
\begin{equation}
\label{6}
[{\mathbf x}^\mu,{\mathbf \theta}^{\alpha\beta}] = 0\,\,.
\end{equation}

With this formalism, DFR demonstrated that after the combination of QM with classical gravitation theory, the ordinary spacetime lost all operational meaning at short distances.

An important point in DFR algebra is that the Weyl representation of NC operators obeying 
(\ref{a1}) and (\ref{6}) keeps the usual form of the Moyal product, 
\begin{equation} \label{a11.22}
\varphi(x)\,\star\,\psi(x)\,=\,exp\Big[\frac{i}{2}\,\theta^{\mu\nu}\,\frac{\partial}{\partial\,x^{\mu}}\frac{\partial}{\partial\,y^{\nu}} \Big]\,\varphi(x)\,\psi(y)\mid_{x=y}\,\,,
\end{equation}
and consequently the form of the usual NCFT's, although the fields have to be considered as dependent not only on ${\mathbf x}^\mu$ but also on ${\mathbf \theta}^{\alpha\beta}$.
The argument is that  very accurate measurements of spacetime localization can transfer to test particles, sufficient energies to create a gravitational field that in principle can trap photons. This possibility is related to spacetime uncertainty relations that can be derived from (\ref{a1}) and (\ref{6}) as well as from the quantum conditions 
\begin{eqnarray}
\label{04000}
&&{\mathbf \theta}_{\mu\nu}{\mathbf \theta}^{\mu\nu} =0\nonumber\\
&&({1\over4}\;{}^*{\mathbf \theta}^{\mu\nu}{\mathbf \theta}_{\mu\nu})^2=\lambda_{Pl}^8
\end{eqnarray}

\noindent where $^*{\mathbf \theta}_{\mu\nu}={1\over2}\epsilon_{\mu\nu\rho\sigma}{\mathbf \theta}^{\rho\sigma}$.

In this section we will furnish the main ingredients of the extended DFR algebra.  The interested reader can find more details and an extensive list of other approaches for the NC theories in \cite{QG}.

Remembering that the DFR algebra \cite{DFR} essentially assumes (\ref{a1}) as well as the vanishing of the triple commutator among the coordinate operators,
\be \label{triple}
[\,{\mathbf x}^{\mu}\,,\,[{\mathbf x}^{\nu}\,,\,{\mathbf x}^{\rho}\,]\,]\,=\,0\,\,.
\ee
It is easy to realize that this relation constitute a constraint in a NC spacetime.  Notice that the commutator inside the triple one is not a $c$-number.

The basic DFR algebra relies on  principles imported from GR and QM. In addition to (\ref{a1}) it also assumes that 
\begin{equation}
\label{6a}
[{\mathbf x}^\mu,{\mathbf \theta}^{\alpha\beta}] = 0\,\,,
\end{equation}

\ni and we consider that space has arbitrary $D\geq2$ dimensions.  As usual ${\mathbf x}^\mu$ and ${\mathbf p}_\nu$, where $i,j=1,2,...,D$,  and  $\mu,\nu=0,1,....,D$,  represent the position operator and its
conjugate momentum in Euclidean and Minkowski spaces respectively.   The NC variable ${\mathbf \theta}^{\mu\nu}$ represent the noncommutativity operator, but now ${\mathbf \pi}_{\mu\nu}$ is its conjugate momentum. In accordance with the discussion above, it follows the algebra
\begin{subequations}
\label{7000}
\ba
\big[{\mathbf x}^\mu,{\mathbf p}_\nu \big] &=& i \delta^\mu_\nu\,\,,\label{7000a} \\
\big[\,\theta^{\mu\nu},\pi_{\alpha\beta}\, \big] &=& i \delta^{\mu\nu}_{\,\,\,\,\alpha\beta}\,\,,\label{7000b}
\ea
\end{subequations}

\noindent where $\delta^{\mu\nu}_{\,\,\,\,\alpha\beta}=\delta^{\mu}_{\alpha}\delta^{\nu}_{\beta}-\delta^{\mu}_{\beta}\delta^{\nu}_{\alpha}$. 
The relation (\ref{a1}) here in a space with $D$ dimensions, for example, can be written as
\begin{equation}
\label{9000}
[{\mathbf x}^i,{\mathbf x}^j] = i \,{\mathbf \theta}^{ij} \qquad \mbox{and} \qquad [{\mathbf p}_i,{\mathbf p}_j ] = 0
\end{equation}

\noindent and together with the triple commutator (\ref{triple}) condition for the standard spacetime, i.e., 
\begin{equation}
\label{10}
[{\mathbf x}^\mu,{\mathbf \theta}^{\nu\alpha}] = 0\,\,.
\end{equation}

\noindent This implies that 
\begin{equation}
\label{11}
[{\mathbf \theta}^{\mu\nu},{\mathbf \theta}^{\alpha\beta}] = 0\,\,,
\end{equation} 
and this completes the DFR algebra.

Recently, in order to obtain consistency it was introduced \cite{Amorim1}, as we talked above, the canonical conjugate momenta $\pi_{\mu\nu}$ such that,
\begin{equation}
\label{12}
[{\mathbf p}_\mu,{\mathbf \theta}^{\nu\alpha}] = 0\,\,,\,\,\,\,\,\,
[{\mathbf p}_\mu,{\mathbf \pi}_{\nu\alpha}] = 0\,\,.
\end{equation}

The Jacobi identity formed by the operators ${\mathbf x}^i$, ${\mathbf x}^j$ and ${\mathbf \pi}_{kl}$ leads to the nontrivial relation
\begin{equation}
\label{14}
[[{\mathbf x}^\mu,{\mathbf \pi}_{\alpha\beta}],{\mathbf x}^\nu]- [[{\mathbf x}^\nu,{\mathbf \pi}_{\alpha\beta}],{\mathbf x}^\mu]   =   - \delta^{\mu\nu}_{\,\,\,\,\alpha\beta}\,\,.
\end{equation}

\noindent The solution, not considering trivial terms,  is given by
\begin{equation}
\label{15000}
[{\mathbf x}^\mu,{\mathbf \pi}_{\alpha\beta}]=-{i\over 2}\delta^{\mu\nu}_{\,\,\,\,\alpha\beta}{\mathbf p}_\nu \,\,.
\end{equation}

\noindent It is simple to verify that the whole set of commutation relations listed above is indeed consistent under all possible Jacobi identities. Expression (\ref{15000}) suggests that the  shifted coordinate 
operator \cite{Chaichan,Gamboa,Kokado,Kijanka,Calmet}
\begin{equation}
\label{16000}
{\mathbf X}^\mu\equiv{\mathbf x}^\mu\,+\,{1\over 2}{\mathbf \theta}^{\mu\nu}{\mathbf p}_\nu\,\,,
\end{equation}

\noindent commutes with ${\mathbf \pi}_{kl}$. Actually, (\ref{16000}) also commutes with ${\mathbf \theta}^{kl}$ and $ {\mathbf X}^j $, and satisfies a non trivial commutation relation with  ${\mathbf p}_i$  dependent objects, which could be derived from
\begin{equation}
\label{17000}
[{\mathbf X}^\mu,{\mathbf p}_\nu]=i\delta^\mu_\nu
\end{equation}
\ni and
\begin{equation}
\label{i6}
[{\mathbf X}^\mu,{\mathbf X}^\nu]=0\,\,.
\end{equation}
So we see from these both equation that the shifted coordinated operator (\ref{16000}) allows us to recover the commutativity.  Hence, differently form $\hat{x}^\mu$, we can say that $\hat{X}^\mu$ forms a basis in the Hilbert space.  This fact will be important very soon.

To construct an extended DFR algebra in $(x,\theta)$ space, we can write 

$${\mathbf M}^{\mu\nu}\,=\,{\mathbf X}^\mu{\mathbf p}^\nu\,-\,{\mathbf X}^\nu{\mathbf p}^\mu\,-\,\theta^{\mu\sigma}\,\pi_{\sigma}^{\:\:\nu}\,+\,\theta^{\nu\sigma}\,\pi_{\sigma}^{\:\:\mu}\,\,,$$ where ${\mathbf M}^{\mu\nu}$ are the antisymmetric generators for the Lorentz group.  To construct  $\pi_{\mu\nu}$ we have to obey equations
(\ref{7000b}) and (2.9), obviously.   From (\ref{7000a}) we can write the generators for translations as $$P_\mu = - i \partial_\mu\,\,.$$  With these ingredients it is easy to construct the commutation relations
\begin{eqnarray} \label{ABC}
\left[ {\mathbf P}_\mu , {\mathbf P}_\nu \right] &=& 0 \nonumber \\
\left[ {\mathbf M}_{\mu\nu},{\mathbf P}_{\rho} \right] &=& -\,i\,\big(\eta_{\mu\nu}\,{\mathbf P}_\rho\,-\,\eta_{\mu\rho}\,{\mathbf P}_\nu \big) \\
\left[ {\mathbf M}_{\mu\nu},{\mathbf M}_{\rho\sigma} \right] &=& -\,i\,(\,\eta_{\mu\rho}\,{\mathbf M}_{\nu\sigma}\,-\,\eta_{\mu\sigma}\,{\mathbf M}_{\nu\rho}\,-\,\eta_{\nu\rho}\,{\mathbf M}_{\mu\sigma}\,-\,\eta_{\nu\sigma}\,{\mathbf M}_{\mu\rho}\,)\;\;, \nonumber 
\end{eqnarray}
and we can say that ${\mathbf P}_\mu$ and ${\mathbf M}_{\mu\nu}$ are the generators for the extended DFR algebra.  These relations are important because they are essential for the construction of the Dirac equation in this extended DFR configuration $D=10$ $(x,\theta)$ space \cite{Amorim5}.  It can be shown that  the Clifford algebra structure generated by the 10 generalized Dirac matrices $\Gamma$ relies on these relations \cite{QG}. In \cite{Amorim2}, the reader can find a complete discussion about the symmetries involved in this extended DFR space.

\subsection{The symmetries}

In this little section, for completeness, we will review (\cite{aa} and the references within) the symmetries involved with the DFRA elements.  They will be important when we attack the main issue of this paper which the charged scalar field action and the solution of its equations of motion.

Analyzing the Lorentz symmetry in NCQM following the lines above, once we introduce an appropriate theory, for instance, given by a scalar action.  We know, however, that the  elementary particles are classified according to the eigenvalues of the Casimir operators of the inhomogeneous Lorentz group. Hence, let us extend this approach to the Poincar\'{e} group ${\cal P}$. By considering the operators presented here, we can in principle consider 
\be
\label{4.20linha}
{\mathbf G}={1\over2}\omega_{\mu\nu}{\mathbf M}^{\mu\nu}-a^\mu{\mathbf p}_\mu+{1\over2}b_{\mu\nu}{\mathbf \pi}^{\mu\nu}
\ee 
as the generator of some   group  ${\cal P}'$, which has the Poincar\'{e} group as a subgroup.  By 
defining the dynamical transformation of an arbitrary operator $\mathbf{A}$ in ${\cal H}$ in such a way that $\delta \mathbf{A}\,=\,i\,[\mathbf{A},\mathbf{G}]$ we arrive at the set of transformations,

\begin{eqnarray}\label{i19a}
\delta {\mathbf X}^\mu&=&\omega ^\mu_{\,\,\,\,\nu}{\mathbf X}^\nu+a^\mu\nonumber\\ 
\delta{\mathbf p}_\mu&=&\omega _\mu^{\,\,\,\,\nu}{\mathbf p}_\nu\nonumber\\
\delta{\mathbf \theta}^{\mu\nu}&=&\omega ^\mu_{\,\,\,\,\rho}{\mathbf \theta}^{\rho\nu}+ \omega ^\nu_{\,\,\,\,\rho}{\mathbf \theta}^{\mu\rho}+b^{\mu\nu}\nonumber\\
\delta{\mathbf \pi}_{\mu\nu}&=&\omega _\mu^{\,\,\,\,\rho}{\mathbf \pi}_{\rho\nu}+ \omega _\nu^{\,\,\,\,\rho}{\mathbf \pi}_{\mu\rho}\nonumber\\
\delta {\mathbf M}_1^{\mu\nu}&=&\omega ^\mu_{\,\,\,\,\rho}{\mathbf M}_1^{\rho\nu}+ \omega ^\nu_{\,\,\,\,\rho}{\mathbf M}_1^{\mu\rho}+a^\mu{\mathbf p}^\nu-a^\nu{\mathbf p}^\mu\nonumber\\
\delta {\mathbf M}_2^{\mu\nu}&=&\omega ^\mu_{\,\,\,\,\rho}{\mathbf M}_2^{\rho\nu}+ \omega ^\nu_{\,\,\,\,\rho}{\mathbf M_2}^{\mu\rho}+b^{\mu\rho}{\mathbf \pi}_\rho^{\,\,\,\,\nu}+ b^{\nu\rho}{\mathbf \pi}_{\,\,\,\rho}^{\mu}\nonumber\\
\delta {\mathbf x}^\mu&=&\omega ^\mu_{\,\,\,\,\nu}{\mathbf x}^\nu+a^\mu+{1\over2}b^{\mu\nu}{\mathbf p}_\nu 
\end{eqnarray}

\noindent We observe that there is an unexpected term in the last one of (\ref{i19a}) system. This is a consequence of the coordinate operator in (\ref{16000}), which is a nonlinear combination of operators that act on ${\cal H}_1$ and ${\cal H}_2$.

The action of ${\cal P}'$ over the Hilbert space operators is in some sense equal to the action of the 
Poincar\'{e} group with an additional translation operation on the (${\mathbf \theta}^{\mu\nu}$) sector. 
All its generators close in an algebra under commutation, so ${\cal P}'$ is a well defined group of transformations. As a matter of fact, the commutation of two  transformations closes in the algebra

\begin{equation}
\label{i19aa}
[\delta_2,\delta_1]\,{\mathbf y}=\delta_3\,{\mathbf y}
\end{equation}

\noindent where ${\mathbf y}$ represents any one of the operators appearing in (\ref{i19a}). The parameters composition rule is given by
\begin{eqnarray}
\label{i19b}
&&\omega^\mu_{3\,\,\,\,\nu}=\omega^\mu_{1\,\,\,\,\alpha}\omega^\alpha_{2\,\,\,\,\nu}-\omega^\mu_{2\,\,\,\,\alpha}\omega^\alpha_{1\,\,\,\,\nu}\nonumber\\
&&a_3^\mu=\omega^\mu_{1\,\,\,\nu}a_2^\nu-\omega^\mu_{2\,\,\,\nu}a_1^\nu\\
&&b_3^{\mu\nu}=\omega^\mu_{1\,\,\,\rho}b_2^{\rho\nu}-\omega^\mu_{2\,\,\,\rho}b_1^{\rho\nu}-\omega^\nu_{1\,\,\,\rho}b_2^{\rho\mu}+
\omega^\nu_{2\,\,\,\rho}b_1^{\rho\mu}\,\,. \nonumber
\end{eqnarray}

\ni These symmetries can also be dynamically constructed using alternatively the Lagrangian formalism instead of the Hamiltonian one.  The symmetry explored in the Eqs. (\ref{i19a}) is discussed with more details in \cite{Amorim4}

\bigskip

\subsection{The DFRA operators}

\bigskip

From (\ref{16000}) we obtained the commutators (\ref{17000}) and (\ref{i6}) and we recovered the commutativity.  So, it is easy to show that the shifted coordinate operator $X^\mu$ given in (\ref{16000}) also commutes with $\theta^{\mu\nu}$ and $\pi_{\mu\nu}$, hence it can be used to construct a basis for this extended spacetime.

However, differently from \cite{Amorim1} we will use the CCR algebra (last section) to guide us to a mathematically rigorous manner to construct the DFRA operators.

Let us consider (using the CCR algebra described in the last section), for example, a $D=3$ case.  In this case the anti-symmetric tensor $\theta^{ij}$ corresponds to a (pseudo) vector ${\mathbf{\sigma}}$ such that,

\be
\label{oper1}
\theta^{ij}\,=\,\epsilon^{ijk}\,\sigma_k\qquad \Rightarrow \qquad \sigma_k\,=\,{1\over 2} \,\epsilon_{kjl}\,\theta^{jl}
\ee
where $\epsilon^{ijk}$ is the three dimensional Levi-Civitta tensor.

Analogously there is a vector $\mathbf{\tau}$ that corresponds to $\pi_{ij}$ in the form

\be
\label{oper2}
\pi_{ij}\,=\,\epsilon_{ijk}\,\tau^k\qquad \Rightarrow \qquad \tau^k\,=\,{1\over 2} \,\epsilon^{kjl}\,\pi_{jl}
\ee

With both $\mathbf{\sigma}$ and $\mathbf{\tau}$ we can rewrite the relation (\ref{7000b}) such that,

\be
\label{oper3}
[\,\sigma_k\,,\,\tau^l\,]\,=\,i\,\delta^l_k\,\,,
\ee
and this last equation together with Eqs. (\ref{6a})-(\ref{12}) and (\ref{17000}) satisfy the CCR relations seen in the last section.

Following the Stone-von Neumann theorems (last section), every irreducible regular representation is unitarily equivalent to the Schr\"odinger representation as we saw in (\ref{E2}).   Then, in six dimensions we have that,

\be
\label{oper4}
{\cal H} \,=\,(\,\mathbb{R}^6,d^3\mathbf{X}\,d^3\mathbf{\sigma}\,)
\ee
where the operators $\hat{X}^k$ and $\hat{\sigma}_l$ are multiplicative and $\hat{p}_k$ 
and $\hat{\tau}^l$ are partial derivatives, as we saw in Eqs. (\ref{F}) and (\ref{G}).  So, for $\hat{\tau}^k$ we can write,

\be
\label{oper5}
\hat{\tau}^k\,=\,-\,i\,\frac{\partial}{\partial\,\sigma_k}\,\,.
\ee

If we consider the transformation of variables as the one in (\ref{F}),
\be
\label{oper6}
\{\sigma_k\,,\,k=1,2,3\} \qquad \rightarrow \qquad \{\theta^{kl}\,,\,k\,<\,l\,\}
\ee
we have that in a NC space
\be
\label{oper7}
{\cal H} \,=\,(\,\mathbb{R}^{6},d^{3}\mathbf{X}\,d\theta^{12}\,d\theta^{23}\,d\theta^{31}\,)\,\,,
\ee
and the $\hat{\theta}^{kl}$ operators are multiplicative and the $\hat{\pi}_{kl}$ operators can be written as

\ba
\label{oper8}
\hat{\pi}_{kl} &=& \epsilon_{klm}\,\hat{\tau}^m \,=\,-\,i\,\epsilon_{klm}\,\frac{\partial}{\partial \sigma_m} \nonumber \\
&=& -i\,\epsilon_{klm}\sum_{r<s}\,\frac{\partial \theta^{rs}}{\partial \sigma_m}\,\frac{\partial}{\partial \theta^{rs}} \nonumber \\
&=& - {i\over 2}\,\epsilon_{klm}\sum_{r,s}\,\epsilon^{rsm}\,\frac{\partial}{\partial \theta^{rs}} \nonumber \\
&=&- {i\over 2}\,\delta^{rs}_{\:\:\:\:kl}\,\frac{\partial}{\partial \theta^{rs}} \nonumber \\
&=&-\,i\,\frac{\partial}{\partial \theta^{kl}}\,\,,
\ea

\noindent which obeys relation (\ref{15000}).  In an analogous way we can calculate the other operators and we can extend this calculation to spacetime.  Notice that in \cite{Amorim1} although correct, there was not a precise demonstration of the form of operator $\hat{\pi}_{\mu\nu}$.  Besides, we show here that the DFRA algebra obeys the rules of a CCR algebra.


\section{The action and symmetry relations}
\renewcommand{\theequation}{5.\arabic{equation}}
\setcounter{equation}{0}

In this section we construct the NC Klein-Gordon equation and provide its symmetries \cite{aa}.
An important point is that, due to (\ref{a1}), the operator ${\mathbf x}^\mu$ can not
be used to label a possible basis in ${\cal H}$. However, as the components of
${\mathbf X}^\mu$ commute (it can be verified from (\ref{6a})-(\ref{15000}) and  (\ref{16000})), their
eigenvalues  can be used for such purpose. From now on let us denote  by $x$ and
$\theta$ the eigenvalues of ${\mathbf X}$ and ${\mathbf\theta}$. In  \cite{Amorim4}
we have considered these points with some detail and have proposed a way for
constructing  actions representing possible field theories in this extended
$x+\theta$ spacetime.  One of such actions, generalized in order to allow the
scalar fields to be complex, is given by

\begin{equation}
\label{b8}
S=-\int d^{4}\,x\,d^{6}\theta\,\, \Big\{\,\partial^\mu\phi^*\partial_\mu\phi +
\,{{\lambda^2}\over4}\,\partial^{\mu\nu}\phi^*\partial_{\mu\nu}\phi  
+m^2\,\phi^*\phi\Big\}\,\,,
\end{equation}

\noindent where $\lambda$ is a parameter with dimension of length, as the Planck
length, which is introduced due to dimensional reasons. Here we are also suppressing
a possible  factor $\Omega(\theta)$ in the measure, which is a scalar weight
function, used in Refs.   
\cite{Carlson,Haghighat,Carone,Ettefaghi,Morita,Saxell}, 
in a NC gauge
theory context,  to make the connection between the $D=4+6$
and the $D=4$ formalisms. Also $\Box= \partial^\mu\partial_\mu  $, with
$\partial_\mu={{\partial}\over{\partial {x}^\mu}}$ and 
$\Box_{\theta}={1\over2}\partial^{\mu\nu}\partial_{\mu\nu}$,
where $\partial_{\mu\nu}={{\partial\,\,\,}\over{\partial {\theta}^{\mu\nu}}}\,\,$ and  
$\eta^{\mu\nu}=diag(-1,1,1,1)$.
\bigskip

The corresponding Euler-Lagrange equation reads

\begin{eqnarray}
\label{c6}
{{\delta S}\over{\delta\phi}}& =& \,(\Box +\lambda^2\Box_\theta- m^2)\phi^* \nonumber\\
& =& \,\,\,0\,\,,
\end{eqnarray}

\noindent with a similar equation of motion for $\phi$. The action (\ref{b8}) is
invariant under the transformation 

\begin{equation}
\label{i19cc}
\delta
\phi=-(a^\mu+\omega^\mu_{\,\,\,\nu}x^\nu)\,\partial_\mu\phi-{1\over2}(b^{\mu\nu}+2\omega^\mu_{\,\,\,\rho}\theta^{\rho\nu})\,\partial_{\mu\nu}\phi\,\,,
\end{equation}

\noindent besides the phase transformation
\begin{equation}
\label{i19dd}
\delta \phi=-i\alpha\,\phi\,\,,
\end{equation}

\noindent with  similar expressions for $\phi^*$, obtained from (\ref{i19cc}) and
(\ref{i19dd}) by complex conjugation. 
 We observe that 
\noindent (\ref{i19a}) closes in an algebra, as in (\ref{i19aa}), with the same
composition rule defined in (\ref{i19b}). That equation defines  how a complex
scalar field transforms in the $x+\theta$ space under ${\cal P}'$. The
transformation subalgebra generated by (\ref{i19a}) is of course Abelian, although
it could be directly generalized to a more general setting.

Associated with those symmetry transformations, we can define the conserved
currents \cite{Amorim4} 

\begin{eqnarray}
\label{92}
& & j^\mu={{\partial {\cal L}}\over{\partial\partial_\mu\phi}}\delta\phi+
\delta\phi^*{{\partial {\cal L}}\over{\partial\partial_\mu\phi^*}}+
{\cal L}\delta x^\mu\,\,,\nonumber\\
& & j^{\mu\nu}={{\partial {\cal L}}\over{\partial\partial_{\mu\nu}\phi}}\delta\phi+
\delta\phi^*{{\partial {\cal L}}\over{\partial\partial_{\mu\nu}\phi^*}}
+{\cal L}\delta \theta^{\mu\nu}\,\,.
\end{eqnarray}

\noindent Actually, by using Eqs. (\ref{i19a}) we can show, after some algebra, that 

\begin{equation}\label{94}
\partial_\mu j^\mu+\partial_{\mu\nu}j^{\mu\nu}=-{{\delta
S}\over{\delta\phi}}\delta\phi-\delta\phi^*{{\delta S}\over{\delta\phi^*}}\,\,.
\end{equation}

Similar calculations can be found, for instance, in \cite{Amorim4}.
The expressions above allow us to derive a specific charge 
\begin{equation}
\label{95}
Q=-\int d^3 x d^6 \theta\, j^0\,\,,
\end{equation}

\noindent  for each kind of conserved symmetry encoded in (\ref{i19a}), since
\begin{equation}
\label{96}
\dot Q=\int d^3 x d^6 \theta\, (\partial_i j^i+{1\over2}\partial_{\mu\nu}j^{\mu\nu}) 
\end{equation}

\noindent vanishes as a consequence of the divergence theorem in this $x+\theta$ extended space. Let us consider each
specific symmetry  in (\ref{i19a}). For usual $x$-translations, we can write
$j^0=j^0_\mu a^\mu$, permitting us to define the total momentum

\begin{eqnarray}
\label{95.1}
P_\mu& =& -\int d^3 x d^6 \theta\, j^0_\mu\nonumber\\
& =& \int d^3 x d^6
\theta\,(\dot\phi^{*}\partial_\mu\phi+\dot\phi\partial_\mu\phi^{*}-{\cal
L}\delta^0_\mu)\,\,.
\end{eqnarray}

\noindent For $\theta$-translations, we can write that $j^0=j^0_{\mu\nu}b^{\mu\nu}$ and consequently
giving
\begin{eqnarray}
\label{96.1}
P_{\mu\nu}& =& -\int d^3 x d^6 \theta\, j^0_{\mu\nu}\nonumber\\
& =& {1\over2}\int d^3 x d^6
\theta\,(\dot\phi^*\partial_{\mu\nu}\phi+\dot\phi\partial_{\mu\nu}\phi^*)\,\,.
\end{eqnarray}
 
\noindent In a similar way we define the Lorentz charge. By using the operator
\begin{equation}
\label{97}
\Delta_{\mu\nu}=x_{[\mu}\partial_{\nu]}+\theta_{[\mu}^{\,\,\,\alpha}\partial_{\nu]\alpha}\,\,,
\end{equation}

\noindent and defining $j^0={\bar j}^0_{\mu\nu}\omega^{\mu\nu}$, we can write
\begin{eqnarray}
\label{98}
M_{\mu\nu}& =& -\int d^3 x d^6 \theta\, {\bar j}^0_{\mu\nu}\nonumber\\
& =& \int d^3 x d^6
\theta\,(\dot\phi^*\Delta_{\nu\mu}\phi+\dot\phi\Delta_{\nu\mu}\phi^*-{\cal
L}\delta^0_{[\mu}x_{\nu]})\,\,.
\end{eqnarray}

Finally, for the symmetry given by (\ref{i19a}), we obtain the $U(1)$ charge 
\begin{equation}
\label{99}
{\cal Q}= i\,\int d^3 x d^6 \theta\,(\dot\phi^*\phi-\dot\phi\phi^*)\,\,.
\end{equation}

Now let us show that these charges generate the appropriate field transformations 
(and dynamics) in a quantum scenario, as generalized Heisenberg relations.
To start the quantization of such theory, we can define as usual the field momenta
\begin{eqnarray}
\label{100}
& & \pi={{\partial{\cal L}}\over{\partial\dot\phi}}=\dot\phi^*\,\,,\nonumber\\
& & \pi^*={{\partial{\cal L}}\over{\partial\dot\phi^*}}=\dot\phi\,\,,
\end{eqnarray}

\noindent satisfying the non vanishing equal time commutators (in what follows the
commutators are to be understood as equal time commutators)
\begin{eqnarray}
\label{101}
& & [\pi(x,\theta),\phi(x',\theta')]=-i\delta^3(x-x')\delta^6(\theta-\theta')\,\,,\nonumber\\
& & [\pi^*(x,\theta),\phi^*(x',\theta')]=-i\delta^3(x-x')\delta^6(\theta-\theta')\,\,.
\end{eqnarray}

Let us generalize the usual field theory and rewrite the
charges in Eqs. (\ref{95.1})-(\ref{99}) by eliminating the time derivatives of the fields in
favor of the field momenta. After that we use (\ref{101}) to dynamically generate
the symmetry operations.   Having said that, accordingly to (\ref{95.1}) and (\ref{100}),
the spatial translation is generated by
\begin{equation}
\label{102}
P_i=\int d^3 x d^6 \theta\,\Big(\pi\partial_i\phi+\pi^*\partial_i\phi^*\Big)\,\,,
\end{equation}

\noindent and it is trivial to verify, by using (\ref{101}), that 
\begin{equation}
\label{103}
[P_i,{\cal Y}(x,\theta)]=-i\partial_i{\cal Y}(x,\theta)\,\,,
\end{equation}

\noindent where ${\cal Y}$ represents $\phi,\,\phi^*,\pi$ or $\pi^*$. 
The dynamics is generated by 
\begin{equation}
\label{102.1}
P_0=\int d^3 x d^6
\theta\,\Big(\pi^*\pi+\partial^i\phi^*\partial^i\phi+{{\lambda^2}\over4}\partial^{\mu\nu}\phi^*\partial_{\mu\nu}\phi+m^2\phi^*\phi\Big)
\end{equation}

\noindent accordingly to (\ref{95.1}) and (\ref{100}). At this stage it is convenient to
assume that classically $\partial^{\mu\nu}\phi^*\partial_{\mu\nu}\phi\geq0$ to
assure that the Hamiltonian $H=P_0$ is positive definite.  By using the fundamental
commutators (\ref{101}), the equations of motion (\ref{c6}) and the definitions
(\ref{100}), it is possible to prove the Heisenberg relation
\begin{equation}
\label{103.1}
[P_0,{\cal Y}(x,\theta)]=-i\partial_0{\cal Y}(x,\theta)\,\,.
\end{equation}

The $\theta$-translations, accordingly to (\ref{96.1}) and (\ref{100}), are generated by
\begin{equation}
\label{104}
P_{\mu\nu}=\int d^3 x d^6
\theta\,\Big(\pi\partial_{\mu\nu}\phi+\pi^*\partial_{\mu\nu}\phi^*\Big)\,\,,
\end{equation}

\noindent and one obtains trivially  by (\ref{101}) that
\begin{equation}
\label{105}
[P_{\mu\nu},{\cal Y}(x,\theta)]=-i\partial_{\mu\nu}{\cal Y}(x,\theta)\,\,.
\end{equation}

The Lorentz transformations are generated by (\ref{98}) in a similar way. The spatial
rotations generator are given by
\begin{equation}
\label{106}
M_{ij}=\int d^3 x d^6 \theta\,\Big(\pi\Delta_{ji}\phi+\pi^*\Delta_{ji}\phi^*\Big)\,\,,
\end{equation}

\noindent  while the boosts are generated by
\begin{eqnarray}
\label{107}
M_{0i}& =& {1\over2}\int d^3 x d^6 \theta\,\Big\{\pi^*\pi
x_i-x_0(\pi\partial_i\phi+\pi^*\partial_i\phi^*)\nonumber\\
& +& \pi(2\theta_{[i}^{\,\,\,\gamma}\partial_{0]\gamma}-x_0\partial_i)\phi+\pi^*(2\theta_{[i}^{\,\,\,\gamma}\partial_{0]\gamma}-x_0\partial_i)\phi^*\nonumber\\
& +& (\partial_j\phi^*\partial_j\phi+{{\lambda^2}\over{4}}\partial^{\mu\nu}\phi^*\partial_{\mu\nu}\phi+m^2\phi^*\phi)x_i\Big\}\,\,.
\end{eqnarray}

\noindent As can be verified in a direct way for (\ref{106}) and in a little more
indirect way for (\ref{107}),
\begin{equation}
\label{108}
[M_{\mu\nu},{\cal Y}(x,\theta)]=i\Delta_{\mu\nu}{\cal Y}(x,\theta)\,\,,
\end{equation}

\noindent for any dynamical quantity ${\cal Y}$, where $\Delta_{\mu\nu}$ has been
defined in (\ref{97}). At last
we can rewrite (\ref{99}) as
\begin{equation}
\label{109}
{\cal Q}= i\,\int d^3 x d^6 \theta\,\Big(\pi\phi-\pi^*\phi^*\Big)\,\,,
\end{equation}

\noindent and its conjugate, and similar expressions for $\pi$ and $\pi^*$. So, the $\cal{P}$' and (global) gauge transformations can be generated by the action of the operator (analogously as in (\ref{4.20linha}))

\begin{equation}
\label{110}
{G}={1\over2}\omega_{\mu\nu}{M}^{\mu\nu}-a^\mu{P}_\mu+{1\over2}b^{\mu\nu}{P}_{\mu\nu}-\alpha {\cal Q}
\end{equation}

\noindent over the complex fields and their momenta, by using the canonical commutation relations (\ref{101}). In this way the $\cal{P}$' and gauge transformations are generated as generalized Heisenberg relations. 
This result shows the consistence of the above formalism. Furthermore, there are also four Casimir operators defined with the operators given above, with the same form as those previously defined at a first quantized perspective (see \cite{aa}).
So, the structure displayed above is very similar to the usual one found in ordinary quantum complex scalar fields. We can go one step further, by calculating the solutions of the equations of motion through the Green's functions and hence constructing the basic structure related to free bosonic fields. This is our objective from now on.



\section{Plane waves and Green's functions}
\renewcommand{\theequation}{6.\arabic{equation}}
\setcounter{equation}{0}

From the first principles we can construct the generalized Huygens' principle in NC DFRA spacetime.  It can be described as,
\be
\label{AA}
\psi(\hat{\mathbf x}',\theta',t')\,=\,i\,\int d^3 x\,d^6\theta\,\Omega(\theta)\,G(\hat{\mathbf x}',\theta',t';\hat{\mathbf x},\theta,t)\,\psi(\hat{\mathbf x},\theta,t)
\ee
where $\psi(\hat{\mathbf x}',\theta',t')$ is the wave that arrives at $(\hat{\mathbf x}',\theta')$ at time $t'$.  We can call the quantity 
$G(\hat{\mathbf x}',\theta',t';\hat{\mathbf x},\theta,t)$ as the extended Green's function or the extended propagator.  In Planck scale, it can describe the effect of the $\psi(\hat{\mathbf x},\theta,t)$, which was at point $(\hat{\mathbf x},\theta)$ in the past, at time $t<t'$, on the wave 
$\psi(\hat{\mathbf x}',\theta',t')$, which is at point $(\hat{\mathbf x}',\theta')$ at the later time $t'$.  Of course that, if the Green's function 
$G(\hat{\mathbf x}',\theta',t';\hat{\mathbf x},\theta,t)$ is known, we can determine the final physical state $\psi(\hat{\mathbf x}',\theta',t')$, which develops from a given initial state $\psi(\hat{\mathbf x},\theta,t)$, using (\ref{AA}).  With the expression for $G$, we can solve the complete scattering problem.  It is equivalent to the complete solution of Klein-Gordon's equation in NC spacetime, namely, the DFRA spacetime.  Hence, the final form of the Green's function is our cherished result, which will be accomplished completely in the next section.  For now we will describe the basic elements of NC theory.

In order to explore a little more the framework described in the last sections, 
 let us rewrite the generalized charged Klein-Gordon action (\ref{b8}) with source terms as
\ba
\label{0.0}
S\,=\,-\,\int d^4 x d^6 \theta \Big\{\p^\mu\phi^* \p_\mu \phi \,+\,{\lambda^2 \over 4}\,\p^{\mu\nu}\phi^* \p_{\mu\nu} \phi  
\,+\,  \,m^2\,\phi^* \phi \,+\,J^* \phi\,+\,J \phi^* \Big\}\,\,.
\ea
\noindent The corresponding equations of motion   are

\be
\label{0.1}
(\Box +\lambda^2\Box_\theta- m^2)\phi(x,\theta)\,=\,J(x,\theta)
\ee

\noindent as well as its complex conjugate one.
 We have the following formal solution

\be
\label{0.2.1}
\phi(x,\theta)\,=\,\phi_{J=0}(x,\theta)\,+\,\phi_J (x,\theta)\,
\ee

\noindent where, clearly, $\phi_{J=0}(x,\theta)$ is the source free solution and $\phi_J (x,\theta)$ is the solution with $J\neq 0$. 

The Green's function for (\ref{0.1}) satisfies
\be
\label{0.3}
(\Box +\lambda^2\Box_\theta- m^2)\,G(x-x',\theta-\theta')=\,\delta^{10} (x-x',\theta-\theta')\,\,,
\ee
where $\delta^{10} (x-x',\theta-\theta')=\delta^4 (x-x')\,\delta^6 (\theta-\theta')$ and the Dirac's delta functions can be defined in NC spacetime as
\be
\label{0.9}
\delta^4 (x-x')\,=\,{1\over (2\pi)^4}\,\int d^4 K_{(1)} \,e^{iK_{(1)}\cdot (x-x')}\,\,,
\ee
\be
\label {0.10}
\delta^6 (\theta-\theta')\,=\,{1\over (2\pi)^6}\,\int d^6 K_{(2)} \,e^{iK_{(2)}\cdot (\theta-\theta')}\,\,.
\ee

Now let us define
\be
\label{0.7}
X\,=\,(x^\mu,{1\over \lambda}\,\theta^{\mu\nu})
\ee
\noindent and
\be
\label{0.8}
K\,=\,(K^\mu_{(1)},\lambda\,K^{\mu\nu}_{(2)})\,\,,
\ee

\noindent where $\lambda$ is a parameter that carries the dimension of length, as said before.  From (\ref{0.7}) and (\ref{0.8}) we 
write that $K\cdot X = K_{(1)\mu}\,x^\mu + {1\over 2} K_{(2)\mu\nu}\,\theta^{\mu\nu}$.  The factor ${1\over 2}$
is introduced in order to eliminate repeated terms. In what follows it will also be considered that 
$d^{10}K=d^4 K_{(1)}d^6 K_{(2)}$ and $d^{10}X=d^4 x\,d^6 \theta$.

So, from (\ref{0.1}) and (\ref{0.3}) we formally have that
\be
\label{0.4}
\phi_J (X)\,=\,\int d^{10} X' G(X-X')\,J(X')\,\,.
\ee

To derive an explicit form for the Green's function, let us expand $G(X-X')$ in terms of plane waves. Hence, we can write that,
\be
\label{0.12}
G(X-X')\,=\,{1\over (2\pi)^{10}} \int d^{10}K \;\tilde{G}(K)\,e^{iK\cdot (X-X')}\,\,. 
\ee

\noindent Now, from (\ref{0.3}), (\ref{0.9}), (\ref{0.10}) and (\ref{0.12}) we obtain that,
\ba
\label{0.13}
\Big(\Box +\lambda^2\Box_\theta- m^2\Big)\,\int\,{d^{10}K\over (2\pi)^{10}}\,\tilde{G} (K)\,e^{iK\cdot (x-x')} 
\,=\,\int\,{{d^{10} x}\over (2\pi)^{10}}\,e^{iK\cdot (x-x')}
\ea
giving the solution for $\tilde{G}(K)$ as
\be
\label{0.14}
\tilde{G} (K)\,=\,-\,{1\over {K^2\,+\,m^2}}
\ee
where, from (\ref{0.8}), $K^2 = K_{(1)\mu}\,K_{(1)}^\mu\,+\,{\lambda^2 \over 2} K_{(2)\mu\nu}\,K_{(2)}^{\mu\nu}$.

Substituting (\ref{0.14}) in (\ref{0.12}) we obtain
\ba\label{0.17}
G(x-x',\theta-\theta')  
=\,{1\over (2\pi)^{10}}\int d^{9} K\,\int d\,K^0 {1\over {(K^0)^2\,-\,\omega^2}}\,e^{iK\cdot (x-x')} 
\ea
where the NC ``frequency" in the $(x+\theta)$ space is defined to be
\ba
\label{0.18}
\omega\,&=&\,\omega(\vec{K}_{(1)},K_{(2)}) \\
\,&=&\,\sqrt{\vec{K}_{(1)}\cdot \vec{K}_{(1)}\,+\,{\lambda^2\over 2}K_{(2)\mu\nu}\,K_{(2)}^{\mu\nu}-m^2} \nonumber
\ea
which can be understood as the dispersion relation in this DFRA spacetime.  We can see also,
from (\ref{0.17}), that there are two poles $K^0 = \pm\;\omega$ in this framework. Of course
 we can construct an analogous solution for 
$\phi^*_{J} (x,\theta)$.

In general, the poles of the Green's function can be interpreted as masses for the stable particles described by the theory.  We can see directly from equation (\ref{0.18}) that the plane waves in the $(x+\theta)$ space
establish the interaction between the currents in this space and have energy given by $\omega(\vec{K}_{(1)},K_{(2)})$
since $\omega^2=\vec{K}_{(1)}^2+{\lambda^2 \over 2}K_{(2)}^2\,+m^2=K^2_{(1,2)}+m^2$, where
$
K^2_{(1,2)}\,=\,\vec{K}_{(1)}^2+{\lambda^2 \over 2}K_{(2)}^2\,\,.
$
So, one can say that the plane waves that mediate the interaction describe the propagation of  particles in a $x+\theta$ spacetime with a mass 
equal to $m$.
Accordingly to the point described in section {\bf III},  we can assume that the Hamiltonian is positive definite and it is directly related to the hypothesis that 
 $K^2_{(1,2)} =-m^2 <0$. However if  the observables are constrained to a four dimensional spacetime, due to some kind of compactification, the physical mass can have contributions from the NC sector. 
Substituting (\ref{0.4}) and (\ref{0.17}) into (\ref{0.0}), we arrive at the effective action

\ba
\label{0.32}
S_{eff}\,=\,-\,\int\,d^4 x\, d^6 \theta\, d^4 x'\, d^6 \theta'\, J^* (X)\,\int \frac{d^9 K}{(2\pi)^{10}} \int\,dK^0
\frac{1}{(K^{0})^2\,-\,\omega^2\,+\,i\varepsilon}\,e^{iK\cdot(X-X')}\,J(X')\,\,, \nonumber \\
\mbox{}
\ea
which could be obtained, in a functional formalism, after integrating out the fields, which will be calculated in the following sections.  

In the next section we will focus on the construction of the retarded and advanced Green functions in DFRA spacetime.



\section{The NC Retarded and advanced Green functions}
\renewcommand{\theequation}{7.\arabic{equation}}
\setcounter{equation}{0}

Notice that our objective in this work, besides to compute the retarded and advanced Green function  is to analyze also the causality in the DFRA NC spacetime.  Firstly, in order to obtain the solution of equation (\ref{0.1}) and its complex conjugate one, namely,

\begin{subequations}
\label{vi}
\ba
\Big( \Box\,+\, \lambda^2\, \Box_{\theta}\,-\,m^2\,\Big)\,\phi(x,\theta)\,&=&\,J(x,\theta) \label{via} \\
\Big( \Box\,+\, \lambda^2\, \Box_{\theta}\,-\,m^2\,\Big)\,\phi^{*}(x,\theta)\,&=&\,J^{*}(x,\theta) \label{vib}
\ea
\end{subequations}

\noindent These solutions gives us the ``in" and ``out" states respectively given by

\ba
\label{vic}
|\,\alpha_{\theta}\,>\,&=&\,|\,\alpha_{\theta},\,t\rightarrow - \infty\,> \\
|\,\beta_{\theta}\,>\,&=&\,|\,\beta_{\theta},\,t\rightarrow  \infty\,> \\
\ea
where $\alpha_{\theta}$ and $\beta_{\theta}$ are the states in a NC spacetime (DFRA) defined asymptotically, at times 
$t\rightarrow -\infty$ and $t\rightarrow \infty$ respectively.

The solution given in Eq. (\ref{via}) can be written explicitly in terms of the retarded and advanced Green functions as

\be
\label{vie}
\phi(x,\theta)\,=\,\phi_{in}(x,\theta)\,+\,\int\,dy\,G^{(+)}(x-x',\theta-\theta')\,\Big( \Box\,+\, \lambda^2\, \Box_{\theta}\,-\,m^2\,\Big)\,\phi(x',\theta')
\ee

\noindent and

\be
\label{vif}
\phi(x,\theta)\,=\,\phi_{out}(x,\theta)\,+\,\int\,dy\,G^{(-)}(x-x',\theta-\theta')\,\Big( \Box\,+\, \lambda^2\, \Box_{\theta}\,-\,m^2\,\Big)\,\phi(x',\theta')
\ee
where $G^{(+)}(x-x',\theta-\theta')$ and $G^{(-)}(x-x',\theta-\theta')$ are the retarded and advanced Green functions as we will see its calculation in the next section.

With the solutions (\ref{vie}) and (\ref{vif}) we can construct an expression for the operator $S$ in DFRA spacetime like,
\be
\label{vig}
\phi_{out}(x,\theta)\,=\,S^{\dagger}\,\phi_{in}(x,\theta)\,S\,\,.
\ee

\noindent One can ask if the LSZ condition (weak asymptotic condition) is valid in DFRA spacetime.  An expression like

\be
\label{vih}
\displaystyle\lim_{t \rightarrow \pm \infty} < a_{\theta}\,|\,\phi(x,\theta)\,|\,b_{\theta} >\,
=\,< a_{\theta}\,|\,\phi_{\overset{out}{in}}\,(x,\theta)\,|\,b_{\theta} >
\ee
where $|\,a_{\theta}\,>$ and $|\,b_{\theta}\,>$ are arbitrary states in the extended Hilbert space would have to be valid in this DFRA spacetime.  If the LSZ condition in Planck scale at the NC spacetime is true, the next step is to construct the vacuum expectation value.  But it is beyond the scope of this work.

We start our calculations by considering the Fourier transform of the Green function defined in (\ref{0.32}),
in $(4+6)D$
\begin{eqnarray}\label{IntFourier10D}
G^{(\gamma)}(x-x^{\prime};\theta-\theta^{\prime})
=\int \frac{d^3{\bf k}}{(2\pi)^3}
\frac{d^3{\bf q}}{(2\pi)^3}
\frac{d^3 \widetilde{{\bf q}}}{(2\pi)^3}
\int_{\gamma}
\frac{d\omega  }{2\pi} \frac{e^{-iK\cdot(X-X^{\prime})}}
{\omega^{2}-{\bf k}^2-\lambda^{2}({\bf q}^{2}+\widetilde{{\bf q}}^{2})} \; ,
\end{eqnarray}
and using the definitions that we have shown in the earlier section
$K\cdot(X-X^{\prime})=k\cdot(x-x^{\prime})
+{\bf q}\cdot(\theta-\theta^{\prime})
+\widetilde{{\bf q}}\cdot(\widetilde{\theta}-\widetilde{\theta}^{\prime})$,
where $(\widetilde{{\bf q}},\widetilde{\theta})$ are dual of $(q_{ij},\theta_{ij})$.
Those integrals can be written as
\begin{eqnarray}\label{IntFourier10Dqqkomega}
G^{(\gamma)}(x-x^{\prime};\theta-\theta^{\prime})
=\int\; \frac{d^3{\bf k}}{(2\pi)^3} \; e^{i{\bf k}\cdot({\bf x}-{\bf x^{\prime}})}\;
\int\; \frac{d^3{\bf q}}{(2\pi)^3} \; e^{i{\bf q}\cdot({\bf \theta}-{\bf \theta^{\prime}})}
\nonumber \\
\times\int\; \frac{d^3 \widetilde{{\bf q}}}{(2\pi)^3} \; e^{i\widetilde{{\bf q}}\cdot(\widetilde{\theta}-\widetilde{\theta}^{\prime})}
\int_{\gamma}
\frac{d\omega  }{2\pi} \frac{e^{-i\omega  (t-t^{\prime})}}
{\omega^{2}-{\bf k}^2-\lambda^{2}({\bf q}^{2}+\widetilde{{\bf q}}^{2})} \; ,
\end{eqnarray}
where $\gamma$ is a contour in the complex $\omega  $-plane infinitesimally close to the real axis
but bypassing the singularities in this axis. The poles on the complex $\omega  $-plane are given
by
\begin{eqnarray} \label{polosinfinitos}
\omega_{\pm}=\pm\omega({\bf k}) \; ,
\end{eqnarray}
where $\omega({\bf k})=\sqrt{{\bf k}^{2}+\lambda^{2}({\bf q}^{2}+\widetilde{{\bf q}}^{2}})$.
The choice in the Green function
(\ref{IntFourier10Dqqkomega}) of a contour $\gamma$ bypassing the poles at the points
$\omega  =\pm\omega({\bf k})$ on the real axis is equivalent to displace them by infinitesimal imaginary quantities.
The displacement $\pm\omega({\bf k})\mapsto \pm\omega({\bf k})-i\varepsilon$ gives rise to the retarded Green function
of the wave equation and the displacement $\pm\omega({\bf k})\mapsto \pm\omega({\bf k})+i\varepsilon$, to the advanced
Green function of the wave equation. For the so called causal Green function the displacement is given by the Feynman
prescription $\pm\omega({\bf k})\mapsto \pm(\omega({\bf k})-i\varepsilon)$.
These Green functions can be calculated by the usual complex variable  technique of closing the contour
$\gamma$ by an infinity semicircle on the upper or lower half plan, as we will do in the next section.

For the retarded Green function prescription we take the poles at
\begin{eqnarray} \label{polosret}
\omega_{\pm}=\pm\omega({\bf k})-i\varepsilon \; ,
\end{eqnarray}
as illustrated by Figure 1. Let us notice that for convenience the poles originally outside the real
axis have also been displaced by $-i\varepsilon$, which is of no consequence due to the usual limit
$\varepsilon\rightarrow 0$ taken at the end of the calculations.
%
%
%
%
\begin{figure}[!h]
\begin{center}
\newpsobject{showgrid}{psgrid}{subgriddiv=1,griddots=10,gridlabels=6pt}
\begin{pspicture}(0,-4)(1,2.5)
\psset{arrowsize=0.2 2}
%
%
\psline[linecolor=black,linewidth=0.4mm]{->}(0,-2.5)(0,2)
\psline[linecolor=black,linewidth=0.4mm]{->}(-4,0)(4,0)
%
%
%
\psline[linecolor=black,linewidth=0.5mm](-1.5,0)(-1.5,-0.2)
\psline[linecolor=black,linewidth=0.5mm](1.5,0)(1.5,-0.2)
%
%
%
\put(3.8,-0.5){$\Re(\omega)$}
\put(0.3,2){$\Im(\omega)$}
\put(1.6,-0.4){$\omega({\bf k})-i\varepsilon$}
\put(-3.6,-0.4){$-\omega({\bf k})-i\varepsilon$}
%
\put(1.49,-0.2){\circle*{0.1}}
%
\put(-1.5,-0.2){\circle*{0.1}}
%
%
%
\end{pspicture}
\vspace{-1.5cm}
\caption{Poles in the complex $\omega$-plane for the calculation of the retarded Green function.}
\label{contornoporbaixo}
\end{center}
\end{figure}
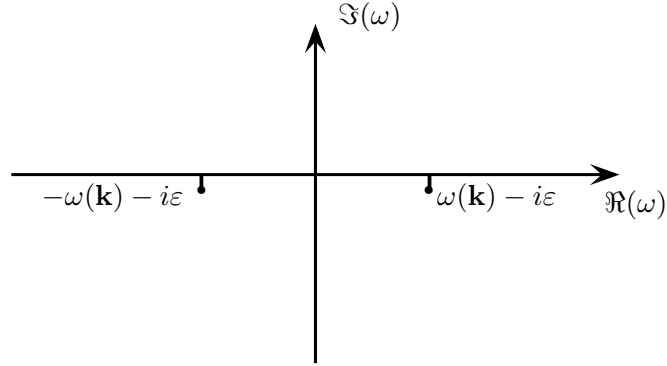
We indicate that the retarded prescription (\ref{polosret})
has been enforced on the Green function (\ref{IntFourier10Dqqkomega}),
by replacing
in it the symbol $(\gamma)$ by $(+)$.
For $t-t^{\prime}>0$ the $\omega$-integration on the real axis is extended
to a closed contour with the usual
infinite radius semicircle on the lower imaginary half-plane.
Applying Cauchy's integral
to those contour integral, we obtain
\begin{eqnarray}\label{IntG+porbaixo}
\left.G^{(+)}(x-x^{\prime};\theta-\theta^{\prime})\right|_{t >t ^{\prime}}
= \int\; \frac{d^3{\bf k}}{(2\pi)^3} \; e^{i{\bf k}\cdot({\bf x}-{\bf x^{\prime}})}\;
\int\; \frac{d^3{\bf q}}{(2\pi)^3} \; e^{i{\bf q}\cdot({\bf \theta}-{\bf \theta^{\prime}})}
\nonumber \\
\times\int\; \frac{d^3 \widetilde{{\bf q}}}{(2\pi)^3} \; e^{i\widetilde{{\bf q}}\cdot(\widetilde{\theta}-\widetilde{\theta}^{\prime})}
\frac{\sin\left[\left(t-t^{\prime}\right)\sqrt{{\bf k}^{2}+\lambda^{2}({\bf q}^{2}+\widetilde{{\bf q}}^{2}}) \right]}{\sqrt{{\bf k}^{2}+\lambda^{2}({\bf q}^{2}+\widetilde{{\bf q}}^{2}})} \; ,
\end{eqnarray}
where the limit $\varepsilon\rightarrow 0$ has already been taken.
Now we perform all the angular integrals in the integration variables $({\bf k},{\bf q},\widetilde{{\bf q}})$,
we obtain the following result
\begin{eqnarray}\label{IntG+dkdqdq}
\left.G^{(+)}(x-x^{\prime};\theta-\theta^{\prime})\right|_{t >t ^{\prime}}
=\frac{1}{|{\bf x}-{\bf x}^{\prime}|}\frac{1}{|{\bf \theta}-{\bf \theta}^{\prime}|}
\frac{1}{|\widetilde{{\bf \theta}}-\widetilde{{\bf \theta}}^{\;\prime}|} \int_{0}^{\infty}
\frac{dk}{(2\pi)^2}  k\sin(k|{\bf x}-{\bf x}^{\prime}|)
\nonumber \\
\times\int_{0}^{\infty} \frac{dq}{2\pi^{2}} q \sin(q|{\bf \theta}-{\bf \theta}^{\prime}|)
\int_{0}^{\infty} \frac{d\widetilde{q}}{2\pi^{2}} \widetilde{q} \sin(\widetilde{q}|\widetilde{\theta}
-\widetilde{ \theta}^{\prime}|)
\frac{\sin\left[\left(t-t^{\prime}\right)\sqrt{k^{2}+\lambda^{2}(q^{2}+\widetilde{q}^{2}}) \right]}{\sqrt{k^{2}+\lambda^{2}(q^{2}+\widetilde{q}^{2}})} \; .
\end{eqnarray}
The integration $(q,\widetilde{q})$ can be performed by using a table of integrals
\cite{Gradshteyn00}, so we will calculate the integrals to obtain
%
%
%
\begin{eqnarray}\label{IntG+dk}
\left.G^{(+)}(x-x^{\prime};\theta-\theta^{\prime})\right|_{t >t ^{\prime}}
=\frac{1}{8\pi(2\pi)^{3}\lambda|{\bf x}-{\bf x}^{\prime}|}
\frac{1}{\left[\lambda^{2}(t-t^{\prime})^{2}-|\theta-\theta^{\prime}|^{2}
-|\widetilde{\theta}-\widetilde{\theta}^{\prime}|^{2} \right]^{3/2}}
\nonumber \\
\left\{\left[ 
-\frac{|\theta-\theta^{\prime}|^{2}
+|\widetilde{\theta}-\widetilde{\theta}^{\prime}|^{2}}{2\lambda^{4}(t-t^{\prime})^{3}}
+\frac{\lambda^{2}(t-t^{\prime})^{2}-|\theta-\theta^{\;\prime}|^{2}
-|\widetilde{\theta}-\widetilde{\theta}^{\prime}|^{2}}{2\lambda^{4}(t-t^{\prime})^{2}}
\frac{d}{dt^{\prime}}\right]
\right. \nonumber \\
\left. \left[\frac{\delta^{\prime}(t^{\prime}-t_{r})}{\sqrt{|{\bf x}-{\bf x}^{\prime}|^{2}
+\lambda^{-2}|\theta-\theta^{\prime}|^{2}
+\lambda^{-2}|\widetilde{\theta}-\widetilde{\theta}^{\prime}|^{2}}}
-\frac{\delta^{\prime}(t^{\prime}-t_{r})|{\bf x}-{\bf x}^{\prime}|^{2}}{\left(|{\bf x}-{\bf x}^{\prime}|^{2}
+\lambda^{-2}|\theta-\theta^{\prime}|^{2}
+\lambda^{-2}|\widetilde{\theta}-\widetilde{\theta}^{\prime}|^{2}\right)^{3/2}}
\right. \right. \nonumber \\
\left. \left.
+\frac{\delta^{\prime \prime}(t^{\prime}-t_{r})|{\bf x}-{\bf x}^{\prime}|^{2}}{|{\bf x}-{\bf x}^{\prime}|^{2}
+\lambda^{-2}|\theta-\theta^{\prime}|^{2}
+\lambda^{-2}|\widetilde{\theta}-\widetilde{\theta}^{\prime}|^{2}}\right]
\right. \nonumber \\
\left. -\frac{1}{t-t^{\prime}}\left[\frac{\delta^{\prime}(t^{\prime}-t_{r})}{\sqrt{|{\bf x}-{\bf x}^{\prime}|^{2}
+\lambda^{-2}|\theta-\theta^{\prime}|^{2}
+\lambda^{-2}|\widetilde{\theta}-\widetilde{\theta}^{\prime}|^{2}}}
-\frac{\delta^{\prime}(t^{\prime}-t_{r})|{\bf x}-{\bf x}^{\prime}|^{2}}{\left(|{\bf x}-{\bf x}^{\prime}|^{2}
+\lambda^{-2}|\theta-\theta^{\prime}|^{2}
+\lambda^{-2}|\widetilde{\theta}-\widetilde{\theta}^{\prime}|^{2}\right)^{3/2}}
\right. \right. \nonumber \\
\left. \left. +\frac{\delta^{\prime\prime}(t^{\prime}-t_{r})|{\bf x}-{\bf x}^{\prime}|^{2}}{|{\bf x}-{\bf x}^{\prime}|^{2}
+\lambda^{-2}|\theta-\theta^{\prime}|^{2}
+\lambda^{-2}|\widetilde{\theta}-\widetilde{\theta}^{\prime}|^{2}} \right]
\right. \nonumber \\
\left.
-\frac{3/2\delta(t^{\prime}-t_{r})}{(t-t^{\prime})^{2}
+\frac{|\theta-\theta^{\;\prime}|^{2}}{\lambda^{2}}
+\frac{|\widetilde{\theta}-\widetilde{\theta}^{\prime}|^{2}}{\lambda^{2}}}
\left[\frac{1}{\sqrt{|{\bf x}-{\bf x}^{\prime}|^{2}
+\lambda^{-2}|\theta-\theta^{\prime}|^{2}
+\lambda^{-2}|\widetilde{\theta}-\widetilde{\theta}^{\prime}|^{2}}}
\right. \right. \nonumber \\
\left. \left.
-\frac{|{\bf x}-{\bf x}^{\prime}|^{2}}{\left(|{\bf x}-{\bf x}^{\prime}|^{2}
+\lambda^{-2}|\theta-\theta^{\prime}|^{2}
+\lambda^{-2}|\widetilde{\theta}-\widetilde{\theta}^{\prime}|^{2}\right)^{3/2}}
+\frac{|{\bf x}-{\bf x}^{\prime}|^{2}}{|{\bf x}-{\bf x}^{\prime}|^{2}
+\lambda^{-2}|\theta-\theta^{\prime}|^{2}
+\lambda^{-2}|\widetilde{\theta}-\widetilde{\theta}^{\prime}|^{2}} \right] \right\} \; , \nonumber \\
\end{eqnarray}
where we have written the delta functions in terms of retarded time $t_{r}$ in $(4+6)D$
that is given by
\begin{eqnarray}\label{tr}
t_{r}=t-\sqrt{{|{\bf x}-{\bf x}^{\prime}|^{2}
+\frac{|\theta-\theta^{\prime}|^{2}}{\lambda^{2}}
+\frac{|\widetilde{\theta}-\widetilde{\theta}^{\prime}|^{2}}{\lambda^{2}}}} \; ,
\end{eqnarray}
and $t-t^{\prime}>0$.

The retarded Green function computed in (\ref{IntG+dk}) allows us to write the asymptotical solutions (\ref{vie}) of the equations (\ref{via}) and (\ref{vib}).  We can see from (\ref{tr}) that the retarded time is affected by the noncommutativity introduced in the Klein-Gordon theory.  From (\ref{IntG+dk}) we realize that the presence of the NC parameter does not constitute a point of singularity.
It is obvious that when these parameters are zero we have the commutative theory.  The advanced Green function is calculated in an analogous way and we will display its form here in order to avoid another big expression like the one in (\ref{IntG+dk}).  It is clear that the same features present in the retarded Green function will analogously appear in the advanced Green function.

With the retarded and advanced NC Green functions we can construct the NC version of the Pauli-Jordan function, i.e.,

\be
\label{vii}
G(x,\theta)\,=\,G^{(+)}(x,\theta)\,-\,G^{(-)}(x,\theta)\,\,,
\ee
which satisfy the homogeneous NC Klein-Gordon equation with $J(x,\theta)=J^* (x,\theta)=0$ in (\ref{via}) and (\ref{vib}) respectively.  Hence, with the solution for the retarded Green function obtained in (\ref{IntG+dk}) we can also calculate the homogeneous solution of equation (\ref{0.1}).

We can also construct the NC Pauli-Jordan function in the following way,
\be
\label{vij}
G(x,\theta)\,=\,G_{(+)}(x,\theta)\,+\,G_{-}(x,\theta)
\ee
where $G_{(+)}$ and $G_{(-)}$ are the parts of the NC Pauli-Jordan with positive and negative frequencies respectively.

In the next section we will analyze the causality problem in DFRA spacetime.  The standard procedure is to compute the causal Green function (the Feynman propagator)  for the Klein-Gordon equation where the source term is a delta function, i.e., $J(x,\theta)=\delta^9\,(x-x',\theta-\theta')$, in (\ref{via}) and (\ref{vib}).  The principle of causality is also valid in a NC  spacetime like the DFRA one.  Even in the NC part of its phase space, we have an absence of faster-than-light transmission of signals.

The NC causal Green function is an essential ingredient for the analysis of interacting field theory in DFRA spacetime by using a kind of NC perturbation theory, for example.


\section{The NC causal Green function}
\renewcommand{\theequation}{8.\arabic{equation}}
\setcounter{equation}{0}

In this section we will calculate the causal Green function, i.e., the Feynman propagator.  Namely, it is a solution of the inhomogeneous NC Klein-Gordon equation containing a delta function as we said above.

As mentioned, its prescription is by substituting
$\pm\omega({\bf k})\longmapsto (\pm\omega({\bf k})-i\varepsilon)$
in the expression of the Fourier transform (\ref{IntG+dkdqdq})
\begin{eqnarray}\label{IntFourier10DqqkomegaF}
G^{(F)}(x-x^{\prime};\theta-\theta^{\prime})
=\int\; \frac{d^3{\bf k}}{(2\pi)^3} \; e^{i{\bf k}\cdot({\bf x}-{\bf x^{\prime}})}\;
\int\; \frac{d^3{\bf q}}{(2\pi)^3} \; e^{i{\bf q}\cdot({\bf \theta}-{\bf \theta^{\prime}})}
\nonumber \\
\times\int\; \frac{d^3 \widetilde{{\bf q}}}{(2\pi)^3} \; e^{i\widetilde{{\bf q}}\cdot(\widetilde{\theta}-\widetilde{\theta}^{\prime})}
\int_{F}
\frac{d\omega  }{2\pi} \frac{e^{-i\omega  (t-t^{\prime})}}
{\omega^{2}-{\bf k}^2-m^{2}-\lambda^{2}({\bf q}^{2}+\widetilde{{\bf q}}^{2})-i\varepsilon} \; ,
\end{eqnarray}
in which the notation $(F)$ indicates the Feynman Green function where the poles are displaced
as
\begin{eqnarray} \label{polosretcausal}
\omega_{\pm}=\pm\omega({\bf k})-i\varepsilon \; ,
\end{eqnarray}
and illustrated by Figure 2.
%
%
%
%
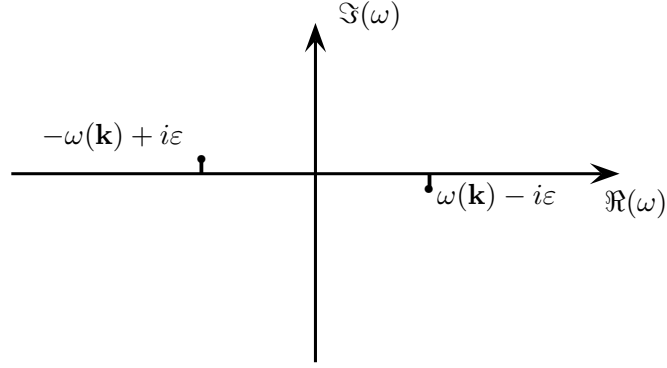
\begin{figure}[!h]
\begin{center}
\newpsobject{showgrid}{psgrid}{subgriddiv=1,griddots=10,gridlabels=6pt}
\begin{pspicture}(0,-4)(1,2.5)
\psset{arrowsize=0.2 2}
%
%
\psline[linecolor=black,linewidth=0.4mm]{->}(0,-2.5)(0,2)
\psline[linecolor=black,linewidth=0.4mm]{->}(-4,0)(4,0)
%
%
%
\psline[linecolor=black,linewidth=0.5mm](-1.5,0)(-1.5,0.2)
\psline[linecolor=black,linewidth=0.5mm](1.5,0)(1.5,-0.2)
%
%
%
\put(3.8,-0.5){$\Re(\omega)$}
\put(0.3,2){$\Im(\omega)$}
\put(1.6,-0.4){$\omega({\bf k})-i\varepsilon$}
\put(-3.6,0.4){$-\omega({\bf k})+i\varepsilon$}
%
\put(1.49,-0.2){\circle*{0.1}}
%
\put(-1.5,0.2){\circle*{0.1}}
%
%
%
\end{pspicture}
\vspace{-1.5cm}
\caption{Poles in the complex $\omega$-plane for the calculation of the causal Green function.}
\label{contornoporbaixocausal}
\end{center}
\end{figure}
The calculation of the $\omega$-integration is also an application
of the Cauchy theorem but with the two contribution of the upper and lower
imaginary half-plane, so we obtain the result
\begin{eqnarray}\label{IntFourier10DqqkF}
G^{(F)}(x-x^{\prime};\theta-\theta^{\prime})
=-i\int\; \frac{d^3{\bf k}}{(2\pi)^3} \; e^{i{\bf k}\cdot({\bf x}-{\bf x^{\prime}})}\;
\int\; \frac{d^3{\bf q}}{(2\pi)^3} \; e^{i{\bf q}\cdot({\bf \theta}-{\bf \theta^{\prime}})}
\nonumber \\
\times\int\; \frac{d^3 \widetilde{{\bf q}}}{(2\pi)^3} \;
\frac{e^{i\widetilde{{\bf q}}\cdot(\widetilde{\theta}-\widetilde{\theta}^{\prime})}}{2\sqrt{{\bf k}^{2}+m^{2}+\lambda^{2}\left({\bf q}^{2}+\widetilde{{\bf q}}^{2} \right)}}
\left[e^{i(t-t^{\prime})\sqrt{{\bf k}^{2}+m^2+\lambda^{2}\left({\bf q}^{2}+\widetilde{{\bf q}}^{2} \right)}}\Theta(t^{\prime}-t)
\right. \nonumber \\
\left. +e^{-i(t-t^{\prime})\sqrt{{\bf k}^{2}+m^{2}+\lambda^{2}\left({\bf q}^{2}+\widetilde{{\bf q}}^{2} \right)}}\Theta(t-t^{\prime}) \right] \; .
\end{eqnarray}
The angular integrals in the variables $({\bf k},{\bf q},\widetilde{{\bf q}})$ are performed to give
\begin{eqnarray}\label{IntGFdkdqdq}
G^{(F)}(x-x^{\prime};\theta-\theta^{\prime})
=-\frac{i}{|{\bf x}-{\bf x}^{\prime}|}\frac{1}{|{\bf \theta}-{\bf \theta}^{\prime}|}
\frac{1}{|\widetilde{{\bf \theta}}-\widetilde{{\bf \theta}}^{\;\prime}|} \int_{0}^{\infty}
\frac{dk}{(2\pi)^2}  k\sin(k|{\bf x}-{\bf x}^{\prime}|)
\nonumber \\
\times\int_{0}^{\infty} \frac{dq}{(2\pi)^{2}} q \sin(q|{\bf \theta}-{\bf \theta}^{\prime}|)
\int_{0}^{\infty} \frac{d\widetilde{q}}{2\pi^{2}} \widetilde{q}
\frac{\sin(\widetilde{q}|\widetilde{\theta}
-\widetilde{ \theta}^{\prime}|)}{\sqrt{k^{2}+m^{2}+\lambda^{2}(q^{2}+\widetilde{q}^{2}})}
\nonumber \\
\times\left[e^{i(t-t^{\prime})\sqrt{k^{2}+m^{2}+\lambda^{2}\left(q^{2}+\widetilde{q}^{2} \right)}}\Theta(t^{\prime}-t) +e^{-i(t-t^{\prime})\sqrt{k^{2}+m^{2}+\lambda^{2}\left(q^{2}+\widetilde{q}^{2} \right)}}\Theta(t-t^{\prime}) \right] \; .
\end{eqnarray}
\begin{eqnarray}\label{IntGFdkdqdqcausal}
G^{(F)}(x-x^{\prime};\theta-\theta^{\prime})
=-\frac{1}{4\pi\lambda|{\bf x}-{\bf x}^{\prime}|}\frac{1}{|{\bf \theta}-{\bf \theta}^{\prime}|}
\frac{1}{\sqrt{(t-t^{\prime})^{2}-|\widetilde{\theta}-\widetilde{\theta}^{\prime}|^{2}}}
\nonumber \\
\times\int_{0}^{\infty}
\frac{dk}{(2\pi)^2}  k\sin(k|{\bf x}-{\bf x}^{\prime}|)
\int_{0}^{\infty} \frac{dq}{(2\pi)^{2}} q \sin(q|{\bf \theta}-{\bf \theta}^{\prime}|)
\sqrt{q^{2}+\frac{k^2+m^{2}}{\lambda^{2}}}
\nonumber \\
\times
\left[H_{1}^{(1)}\left(\sqrt{q^{2}+\frac{k^2+m^{2}}{\lambda^{2}}}\sqrt{(t-t^{\prime})^{2}
-|\widetilde{\theta}-\widetilde{\theta}^{\prime}|^{2}} \right)\Theta(t^{\prime}-t)
\right. \nonumber \\
\left. -H_{1}^{(2)}\left(\sqrt{q^{2}+\frac{k^2+m^{2}}{\lambda^{2}}}\sqrt{(t-t^{\prime})^{2}
-|\widetilde{\theta}-\widetilde{\theta}^{\prime}|^{2}} \right)\Theta(t-t^{\prime}) \right] \; .
\end{eqnarray}
The next integration $(\widetilde{q},q)$ are calculated with help of a integrals handbook
\cite{Gradshteyn00} which result is
\begin{eqnarray}\label{G(F)dk}
G^{(F)}(x-x^{\prime};\theta-\theta^{\prime})
=-\frac{1}{16\pi^{2}\lambda|{\bf x}-{\bf x}^{\prime}|}
\frac{1}{(\lambda^{2}(t-t^{\prime})^{2}-|\theta-\theta^{\prime}|^{2}
-|\widetilde{\theta}-\widetilde{\theta}^{\prime}|^{2})^{3/2}}
\int_{0}^{\infty}
\frac{dk}{(2\pi)^2} k
\hspace{0.5cm} \nonumber \\
\times\sin(k|{\bf x}-{\bf
x}^{\prime}|)\left\{e^{i\sqrt{k^{2}+m^{2}}\sqrt{(t-t^{\prime})^{2}-\frac{|\theta-\theta^{\prime}|^{2}}{\lambda^{2}}
-\frac{|\widetilde{\theta}-\widetilde{\theta}^{\prime}|^{2}}{\lambda^{2}}}}
\left[\frac{k^2+m^{2}}{\lambda^{2}}+\frac{3i\sqrt{k^{2}+m^{2}}}{\lambda\sqrt{\lambda^{2}(t-t^{\prime})^{2}
-|\theta-\theta^{\prime}|^{2}-|\widetilde{\theta}-\widetilde{\theta}^{\prime}|}}
\right. \right. \nonumber \\
\left. \left. -\frac{3}{\lambda^{2}(t-t^{\prime})^{2}
-|\theta-\theta^{\prime}|^{2}-|\widetilde{\theta}-\widetilde{\theta}^{\prime}|} \right]\Theta(t^{\prime}-t)  -e^{-i\sqrt{k^{2}+m^{2}}\sqrt{(t-t^{\prime})^{2}-\frac{|\theta-\theta^{\prime}|^{2}}{\lambda^{2}}
-\frac{|\widetilde{\theta}-\widetilde{\theta}^{\prime}|^{2}}{\lambda^{2}}}}
\hspace{1.5cm} \right. \nonumber \\
\left. \left[\frac{k^2+m^{2}}{\lambda^{2}}-\frac{3i\sqrt{k^{2}+m^{2}}}{\lambda\sqrt{\lambda^{2}(t-t^{\prime})^{2}
-|\theta-\theta^{\prime}|^{2}-|\widetilde{\theta}-\widetilde{\theta}^{\prime}|}}
-\frac{3}{\lambda^{2}(t-t^{\prime})^{2}
-|\theta-\theta^{\prime}|^{2}-|\widetilde{\theta}-\widetilde{\theta}^{\prime}|} \right]\Theta(t-t^{\prime})  \right\} \nonumber \\
\end{eqnarray}

With this object we can define a NC time-ordered product of NC operators.  Notice that in (\ref{G(F)dk}) we had to avoid the singularities in the NC spacetime, i.e., the singularities in the NC phase space where the $\theta$ parameter are coordinates.


\section{Conclusions and perspectives}

To be free of the infinities that dwell in quantum field theory, Snyder described a Lorentz covariant deformation of the Heisenberg algebra.  It has the properties that the position operators are noncommuting and have discrete spectra \cite{Snyder}.  In other words, the ``Snyder space" can be considered a lattice description of space which occurs thanks to the discrete position spectra.

To sum up the procedure, Snyder introduced a five-dimensional spacetime with $SO(4,1)$ as a symmetry group with generators $\mathbf{M}^{AB}$ where $A,B=0,1,2,3,4$.  Moreover, he postulated the identification between coordinates and generators of the $SO(4,1)$ algebra, i.e., $\mathbf{x}^{\mu}=a\,\mathbf{M}^{4\mu}$, where $\mu,\nu=0,1,2,3$ and the parameter $a$ has dimension of length.  In this way he promoted the spacetime coordinates to be Hermitian operators. 
The fact that the open strings theory in the presence of a magnetic background introduced spacetime noncommutativity with the NC parameters $\theta^{\mu\nu}$ being constant violates Lorentz symmetry, but permits one to treat NCFT's as deformations of ordinary quantum field theories.  However, assuming that $\theta^{\mu\nu}$ is in fact a tensor operator with the same hierarchical level as in the $\mathbf{x}$'s could recover the Lorentz invariance.
To treat $\theta^{\mu\nu}$ as an ordinary coordinate of spacetime lead us to construct a canonical conjugate momentum associated to this coordinate.  Hence, all these objects live in a ten dimensional NC spacetime. 
 
After investigating the quantum mechanical consequences of this framework, a NCFT has begun to have its basic elements constructed in this DFRA spacetime.  However, it is not trivial to construct all the operators of the DFRA quantum algebra because we can be led to wrong constructions.  We showed that the CCR algebra and the DFRA one have similarities that must be analyzed.  

To analyze these analogies, we review the algebraic structure, which is compatible with the minimal canonical DFRA algebra.  It is also invariant under an extended Poincar\'e group of symmetries, but keeping, among others, the usual Casimir invariant operators. After that, we demonstrated exactly that the DFRA algebra obeys the same rules of a CCR algebra.  This fact permits us to construct precisely the DFRA operators.  This is a new approach of the DFRA framework.  The DFRA operators appearing in the current literature were constructed in the usual way.  

We have used the complex scalar fields embedded in the DFRA framework where the object of noncommutativity 
$\theta^{\mu\nu}$ represents independent degrees of freedom.   We also have considered arbitrary source terms in order to construct the general solution for the complex scalar fields using the
Green's function technique.   We have provided the whole solution for the NC Klein-Gordon 
equation introduced in \cite{aa}.  Namely, we have calculated the retarded, the advanced and the causal (the Feynman propagator) Green functions in this ten dimensional spaces.  The presence of the NC parameter does not constitute a source of singularities and its elimination recover the commutative results.  With these results we show that other aspects of this new NC quantum field theory in DFRA spacetime can be computed.

The next step in this program, as a future perspective, would be to construct the mode expansion in order to represent the fields in terms of annihilation and creation operators, acting on some Fock space to be properly defined. Also possible compactifications schemes will also be considered. 
These points are under study and will published elsewhere.


\section{Acknowledgments}

EMCA would like to thank professor Jens Mund (Physics Department of Federal University of Juiz de Fora, Minas Gerais, Brazil) for valuable algebraic discussions.



\begin{thebibliography}{30}
\bibitem{NCQM} In Refs.   \cite{Amorim1} and     \cite{Amorim4} it is possible to
find a large amount of references concerning NC quantum mechanics.
\end{thebibliography}
\end{document}